\documentclass[12pt]{article}
\usepackage{geometry}
\usepackage{a4}
\usepackage{graphicx}
\usepackage{epsf}
\usepackage{amsmath}
\usepackage{amssymb}
\usepackage{cite}
\newcommand{\be}{\begin{equation}}
\newcommand{\ee}{\end{equation}}

\newcommand{\Rmnum}[1]{\expandafter\@slowromancap\romannumeral #1@}
\newcommand{\bea}{\begin{eqnarray}}
\newcommand{\eea}{\end{eqnarray}}
\begin{document}
\def\C{{\mathbb{C}}}
\def\R{{\mathbb{R}}}
\def\s{{\mathbb{S}}}
\def\T{{\mathbb{T}}}
\def\Z{{\mathbb{Z}}}
\def\I{{\mathbb{I}}}
\def\W{{\mathbb{W}}}
\def\Bbb{\mathbb}
\def\BZ{\Bbb Z} 
\def\BR{\Bbb R}
\def\BI{\Bbb I}
\def\BW{\Bbb W}
\def\BM{\Bbb M}
\def\BC{\Bbb C} \def\BP{\Bbb P}
\def\CP{\BC\BP}
\begin{titlepage}
\title{Directed Random Walks on Colored, Periodic Lattices: A Gauge Theoretic Approach}
\author{}
\date{
Subhash Mahapatra, Prabwal Phukon, Tapobrata Sarkar
\thanks{\noindent E-mail:~ subhmaha, prabwal, tapo @iitk.ac.in}
\vskip0.4cm
{\sl Department of Physics, \\
Indian Institute of Technology,\\
Kanpur 208016, \\
India}}
\maketitle
\abstract{
\noindent
We define a random walk problem which admits analytic results, on a class of infinite periodic lattices which are directed and colored. 
Our approach is motivated from the fact that such lattices arise in string theoretic constructs of certain gauge theories. 
An operator counting problem in the latter is mapped to a problem in random walks. This is illustrated with several examples.}
\end{titlepage}

Random walks on lattices has been a well studied subject in statistical mechanics over the last few decades and several variants of this problem 
appear in standard textbooks (see, e.g. \cite{rudnick}). 
Solutions of walks on some simple lattices in one and two dimensions offer analytic results (see, e.g \cite{vineyard}), but this is often difficult to achieve in 
more complicated cases. The purpose of this paper is to show that analytic results for a random walk problem defined on a class of infinite periodic lattices  
(whose edges are directed and whose nodes are colored) are obtainable from corresponding gauge theoretic constructions. 

The specific problem that we deal with here is a variant of the counting problem for the number of distinct sites visited in a random walk on a lattice. This has attracted a lot of 
attention since the classic work of \cite{erdos}. For example, a generating function for the second moment of this number was calculated for some examples
in \cite{larralde}. Typically, generating functions for walks and their moments are obtained numerically, see e.g \cite{conway}. It is thus of interest to construct examples where 
analytic control is feasible. 

\begin{figure}[t!]
\begin{minipage}[b]{0.5\linewidth}
\includegraphics[width=2.8in,height=2.3in]{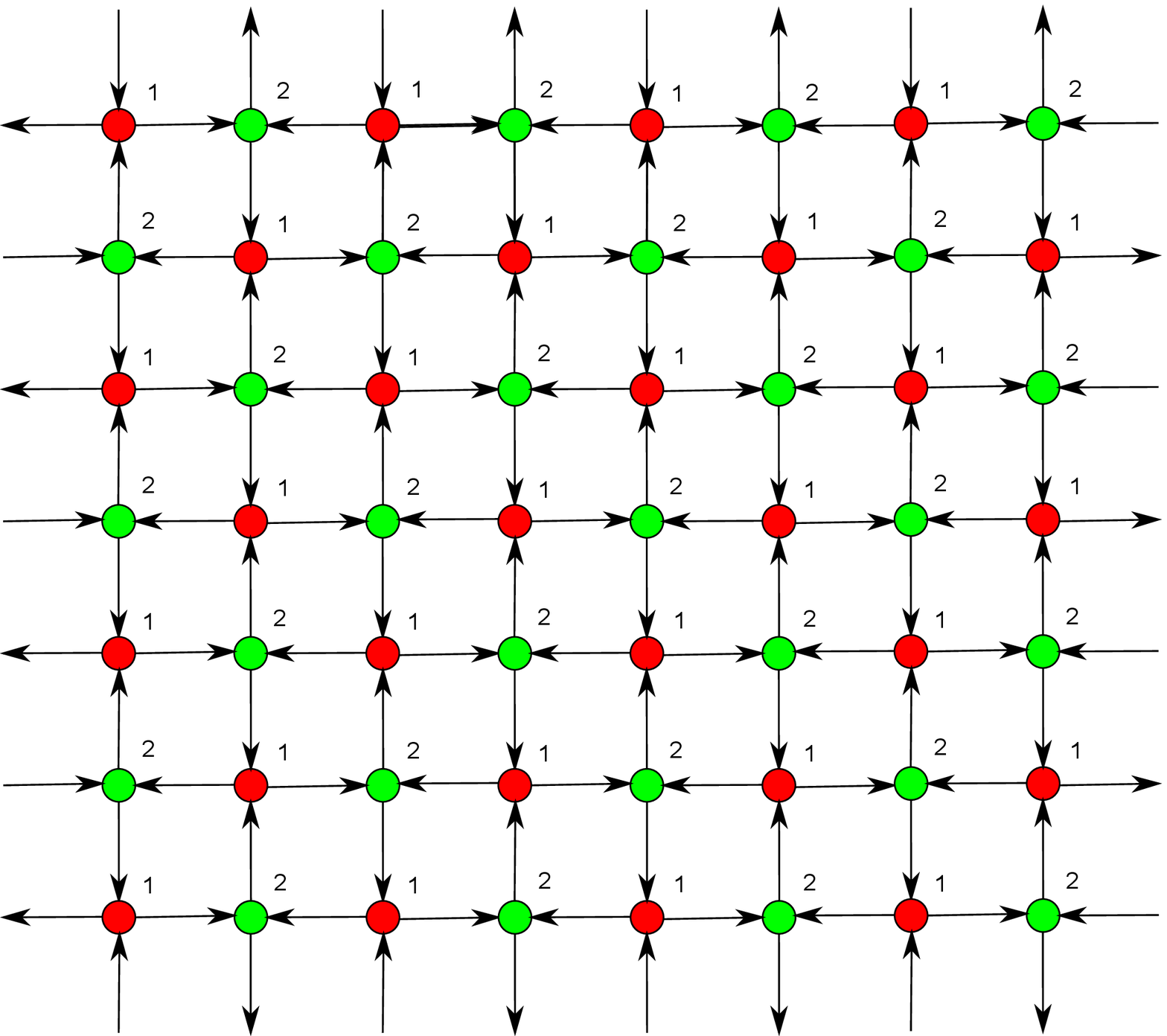}
\caption{An infinite directed square lattice with two colors.}
\label{square1}
\end{minipage}
\hspace{0.2cm}
\begin{minipage}[b]{0.5\linewidth}
\includegraphics[width=2.8in,height=2.3in]{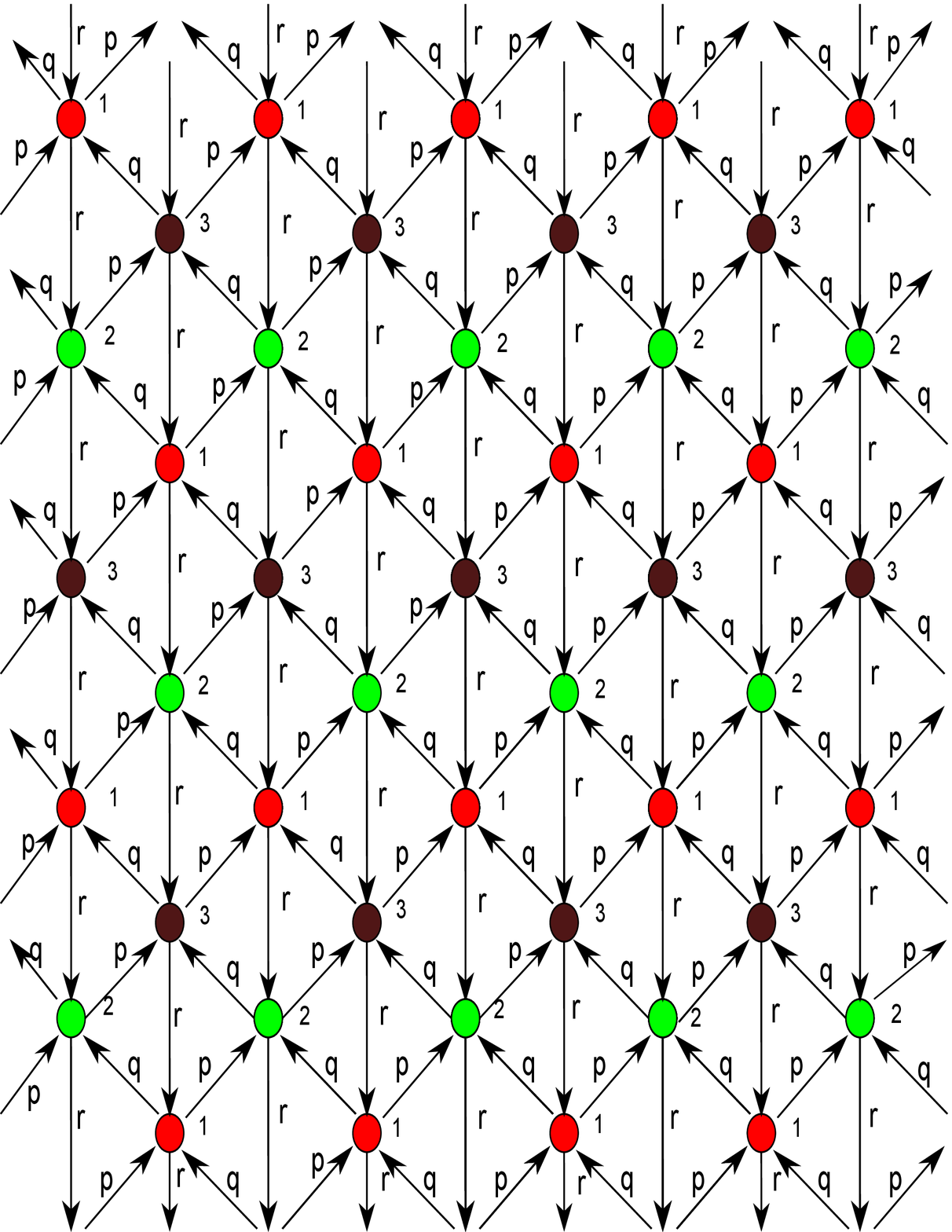}
\caption{An infinite directed triangular lattice with three colors.}
\label{c3z3}
\end{minipage}
\end{figure}

Consider a simple example of an infinite square lattice where the nodes are of two colors, say red and green, and the walks are directed, as shown in fig.(\ref{square1}). Here
the red and green sites have been marked as $1$ and $2$ respectively. The directedness of the lattice implies here that a walker who starts from a red (green) site can go to
his left or right to a green (red) site and is then constrained to move up or down to another red (green) site, and so on. 
Given such a colored, directed lattice, suppose the walker starts from a site of a particular color. We wish to first find out the number of distinct sites {\it of the same color} 
that he visits in an $n$-step walk (which may include the site that he started from). 

The answer is easy to guess in this example : by definition, $n = 2p,~p=0,1,2,\cdots$, is even,  and by inspection, it follows that the number of such sites in a $2p$-step walk 
are the coefficients of $x^{2p}$ in the generating function \footnote{The generating function is denoted by ${\mathcal F}$ throughout, and the subscripts or superscripts 
simply label the cases that we consider.}
\begin{equation}
{\mathcal F}_1^a = \sum_p \left(p+1\right)^2 x^{2p}
\label{squarea}
\end{equation}

Let us reformulate the counting problem slightly. We first define the notion of a ``level n'' walk, where $n = 0,1,2,\cdots$. (This is 
to be distinguished from $n$-step walks \footnote{Such $n$-step walks can be
understood by writing the adjacency matrix for the directed lattice and following the well known methods of \cite{kasteleyn}.}) 
as follows :  start with a lattice site of any particular color, say green. Then, 
a ``level-$1$'' walk is defined to be all possible walks that lead to the site colored green, without crossing a red or green site more than once. This naturally means that 
apart from the beginning and end green sites, the walker does not
come across a green site, in a level-$1$ walk. A level-$2$ walk is now defined as all walks that lead to a green site where the walker does not cross either the red or the
green sites more than twice. This also means that excluding the beginning and end sites, the walker does not cross a green site more than once in a level-$2$ walk. 
This can be generalised naturally to any level-$n$ walk which we are now
ready to define : a level-$n$ 
walk is defined to be one for which the walker, starting from a site of a particular color comes back to a site of the same color without crossing any colored site more than $n$ times 
in-between \footnote{By ``in-between'' we mean excluding the sites that the walker begins from and ends with.}. In the process,
he cannot cross a site having the same color as that of the starting site more than $\left(n-1\right)$ times in-between. It is then simple to check that the number of distinct points 
reached by a level-$n$ walk for the lattice of fig.(\ref{square1}) is given by the coefficients of $x^n$ ($n=0,1,2,\cdots$) in 
\begin{equation}
{\mathcal F}_1^b = \sum_n \left(n+1\right)^2 x^{n}
\label{squareb}
\end{equation}
Thus, we have simply reformulated the counting problem of eq.(\ref{squarea}) \footnote{Note that a level $0$ walk is trivial, as the 
walker remains on his original site.}. 

Now consider a triangular lattice where the infinite, directed lattice sites consist of three colors, say red, green and brown, as shown in fig.(\ref{c3z3}), where we have
also labeled each colored site as $1$, $2$ and $3$ (the letters associated with each of the edges is for future use). The level-$n$ walk can be defined in the same way as alluded 
to before. In fact, the notion of of a level-$n$ walk can be associated to any colored, periodic lattice. 

Suppose we are given an infinite, colored, periodic, directed lattice as the ones just considered. 
We wish to count the number of distinct sites of a particular color visited by a random walker, starting from a given site with the same color, 
in a level-$n$ walk (our definition of a level-$n$ walk incorporates the fact that the walker begins and ends on same-colored sites, and this will be implied in what follows). 
We now argue that in a class of examples, this problem can be mapped to one in gauge theory, and admits an analytic solution. 

Let us start by considering the triangular lattice of fig.(\ref{c3z3}). 
We first state the result (to be discussed momentarily) for its generating function for the number of distinct sites visited by a level-$n$ walk :
\begin{equation}
{\mathcal F}_2 = \frac{1 + 7x + x^2}{\left(1 - x\right)^3}
\label{c3z3series}
\end{equation}
The validity of this result can be seen by expanding the series as 
\begin{equation}
{\mathcal F}_2 = 1 + 10x + 28x^2 + 55x^3 + \cdots
\label{c3z3series1}
\end{equation}
Let our walker start from a green site. A level-$1$ walk will involve three steps, which is the minimum  number of steps required to get back to a green site (apart from
a trivial zero step walk which corresponds to the coefficient $1$ in eq.(\ref{c3z3series1})). For the level-$1$ walk, the number of distinct green sites visited is $10$, as can be
easily seen from fig.(\ref{c3z3}). Similarly, for the level-$2$ walk, the number of distinct green sites visited is $28$ and so on, and the ${\mathcal F}_2$ of eq.(\ref{c3z3series}) 
does indeed provide us with the correct counting. 

We now comment on eq.(\ref{c3z3series}), which has been obtained from a string theoretic perspective. 
In the latter, the two dimensional lattice of fig.(\ref{c3z3}) is obtained as the periodic
``quiver diagram'' for a gauge theory that resides on the four dimensional world volume of a D(irichlet)-brane transverse to the 
resolution of the orbifolded space $\BC^3/\BZ_3$ \cite{dgm}. Here, $\BZ_3$ acts on the coordinates of the three-dimensional complex space $\BC^3$, with
a singular point at the origin.
In the quiver diagram, the nodes denote gauge groups, and the edges represent gauge fields that transform under two such gauge groups. These are 
oppositely charged with respect to the groups under which they transform, and the arrow is conventionally taken from a negative charge to a positive charge. In fig.(\ref{c3z3}) for
example, there are three gauge groups (corresponding to the $\BZ_3$ action on $\BC^3$) which are marked green, red and brown, 
nine distinct gauge fields, with three each transforming simultaneously under 
gauge transformations involving the groups denoted by red and green, red and brown, and green and brown respectively. Eq.(\ref{c3z3series}) in fact counts a particular
class of gauge invariant operators, the ``single trace operators'' for this theory as shown in \cite{hanany}, following the work of \cite{msy}. 
These gauge invariant operators naturally begin and end on the same vertex of the quiver diagram, and hence
conform to our notion of level-$n$ walks (the level-$0$ walk corresponds to the identity operator in the gauge theory). 
We have mapped the counting problem in gauge theory to a random walk problem involving a lattice which is an
infinite version of the quiver diagram for the gauge theory (such infinite quiver diagrams have been considered in \cite{hhv}. 
Indeed, for an infinite number of cases, such a mapping of the gauge theory to a random walk problem is possible. 

The alert reader will also notice that coefficients appearing in the series of eq.(\ref{c3z3series}) represent the Molien series for
the finite group $\BZ_3$, acting on a three dimensional vector space with the generators $\left(1,\omega,\omega^2\right)$, where
$\omega = {\rm e}^{2i\pi/3}$ is a third root of unity. The Molien series counts the number of polynomials a given degree 
that are invariant under the action of the group, $\BZ_3$ in this case. Formally, the Molien series for a discrete group $G$ with order $|G|$ and generators 
$g_i$ is given by
\begin{equation}
{\mathcal M}\left(G,t\right) = \frac{1}{|G|}\sum_{g_i \in G} \frac{1}{{\rm det}\left(\BI - g_i t\right)}
\label{molien}
\end{equation}
This gives the generating function of the number of polynomials invariant under the group action which are the coefficients of various powers of $t$ in eq.(\ref{molien}). 
When calculated for the discrete group $\BZ_3$, this gives the generating function of eq.(\ref{c3z3series}) (with the identification $t^3 = x$).
This can be understood in our context as follows. In fig.(\ref{c3z3}), there are three distinct types of line
orientations, i.e a vertical type, and two other types making angles of $\pi/6$ and $5\pi/6$ to the horizontal. We label each of these by a variable, $\left(r,p,q\right)$ respectively. 
In a three step random walk on this lattice, where we start from a green site and end up on another green site, we collect the integers corresponding to the lines traversed 
by the walker. This gives a distinct collection of polynomials, given by the set
\begin{equation}
S = \left[p^3, q^3, r^3, p^2q, p^2r, pq^2, pr^2, q^2r, qr^2, pqr\right]
\end{equation}
These are the polynomials of order three invariant under the group action $\left(p,q,r\right) \to \omega\left(p,q,r\right)$ where $\omega = {\rm e}^{2i\pi/3}$. 
This correspond to the ten distinct sites visited by our random walker, in a level-$1$ walk. Similar analyses follow for higher level walks. 

\begin{figure}[t!]
\begin{minipage}[b]{0.5\linewidth}
\includegraphics[width=2.8in,height=2.3in]{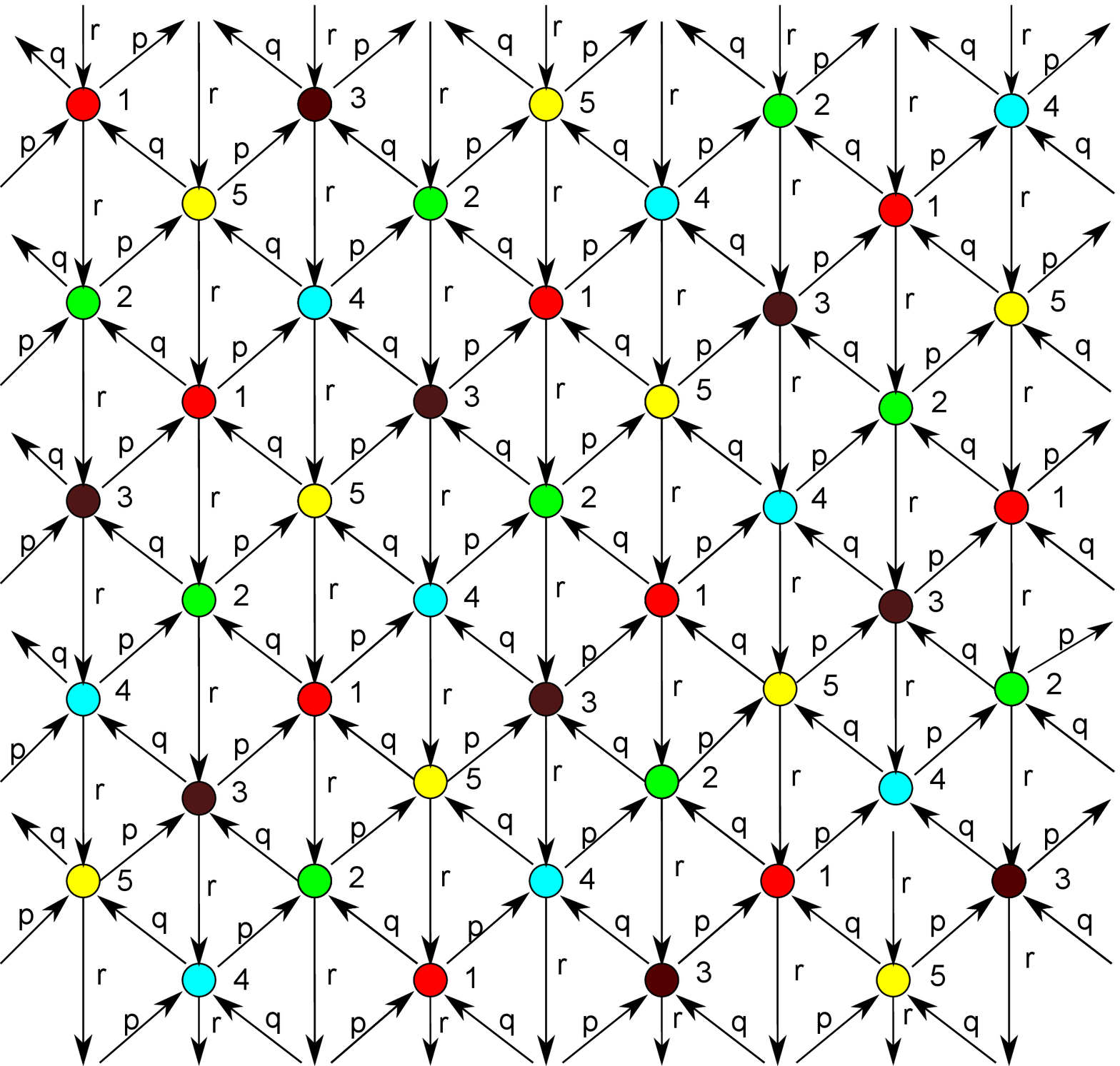}
\caption{Infinite triangular lattice with five colors.}
\label{c3z5}
\end{minipage}
\hspace{0.2cm}
\begin{minipage}[b]{0.5\linewidth}
\includegraphics[width=2.8in,height=2.3in]{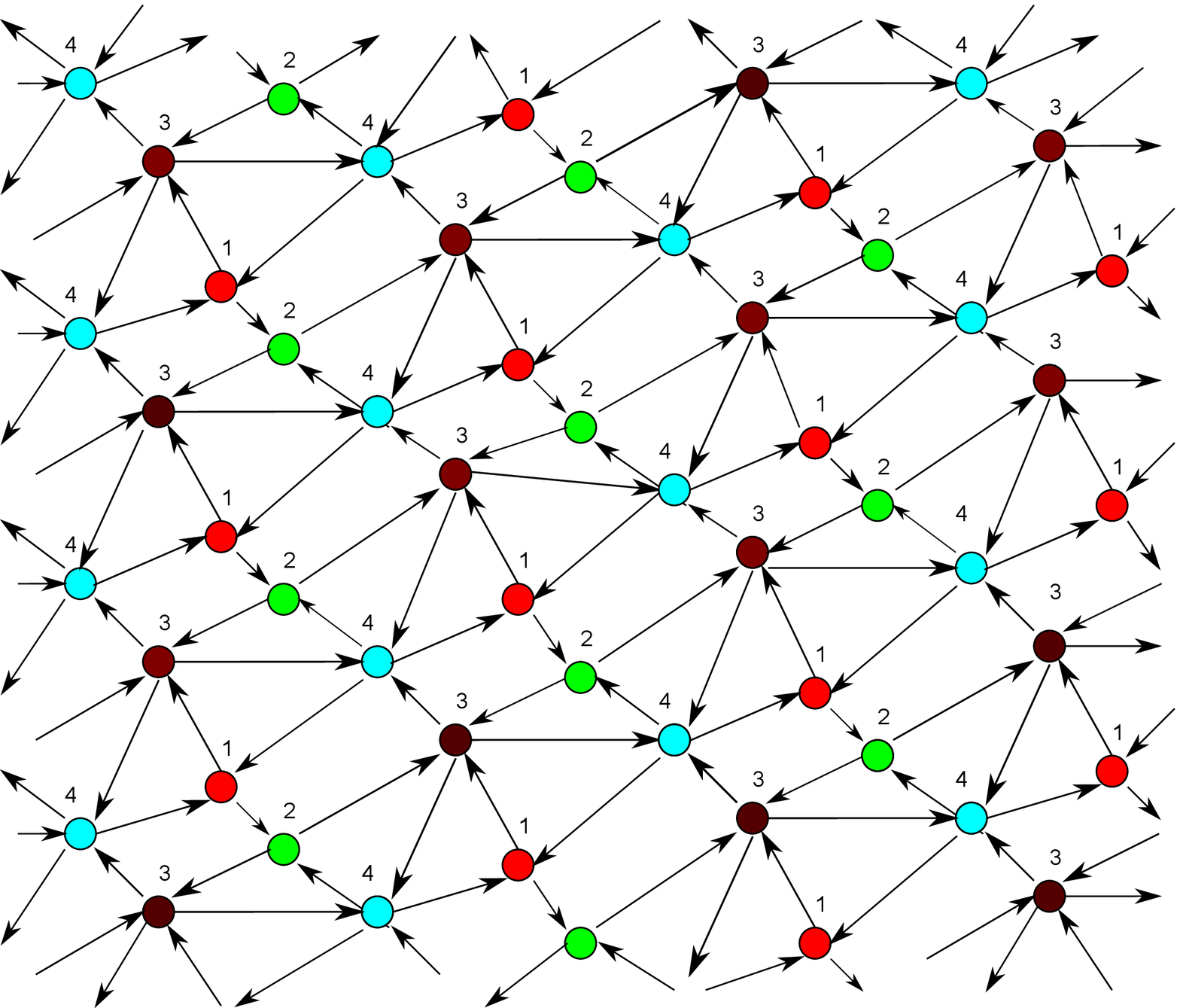}
\caption{An infinite four-colored periodic lattice.}
\label{dP1}
\end{minipage}
\end{figure}

The situation is different if one considers the triangular lattice with five colors, as we have shown in fig.(\ref{c3z5}). Let us first write down the Molien series, which can
be calculated for the group $\BZ_5$ from eq.(\ref{molien}) as 
\begin{equation}
{\mathcal M} = 1 + 3x^3 + 2x^4 + 7x^5 + \cdots
\label{c3z5molien}
\end{equation}
As can be seen, the coefficients of $x^n$ are the number of distinct site of a given color reached while starting from a site of the same color in $n$ steps. The Molien series
is useful for orbifolds of $\BC^3$ by any discrete group $\BZ_n$, $n=3,4,5,\cdots$, and these correspond to random walks on triangular lattices with $n$ colors for the nodes. 
We are however interested in the level-$n$ walks, and the generating function for the level-$n$ walk in this case is given by the the coefficients of $x^n$ in the polynomial
\begin{equation}
{\mathcal F}_3 = \frac{\left(1 + 9x + 5x^2\right)}{\left(1 - x\right)^3} = 1 + 12x + 38x^2 + 79x^3 + \cdots
\label{c3z5series}
\end{equation}
Indeed, it can be checked that starting from a particular site of a given color, there are $12$ distinct sites of the same color that can be reached if none of the other sites
are crossed more than once in-between,  $38$ distinct sites of the same color if none of the other sites are crossed more than twice in-between, and so on.  

For this example, walks starting and ending on the same color 
are possible for $n=0,3,4,\cdots$ steps. For example, for the trivial $0$-step walk, the number of distinct points reached is $1$, for
a $3$-step walk, the walker reaches 3 distinct sites of the same color, and so on. These are the coefficients in the Molien series of eq.(\ref{c3z5molien}). The
level-$1$ walk turns out to be the set of all $3$-, $4$-, and $5$-step walks, and hence, for example, the number of distinct points of the same color reached 
in a level-$1$ walk is $12$, as in eq.(\ref{c3z5series}). Continuing in this way, we find that level-$2$ walks consist of $6$, $7$, $8$ and some of the $9$ and $10$ step walks. The rest of 
the $9$ and $10$ step walks fall into the level-$3$ category.

For the sake of completeness, we now present the recursion relations for the number of distinct sites visited in a level-$n$ walk for the three and five colored triangular 
lattices of figs.(\ref{c3z3}) and
(\ref{c3z5}). Denoting the number of distinct points visited in a level-$n$ walk by $M_n$, we have for the three colored triangular lattice the relation
\begin{equation}
M_{n} = M_{n - 1} + 9n,~~~~n=1,2,3\cdots
\end{equation}
and for the five colored triangular lattice, we get
\begin{equation}
M_n = M_{n-1} + 15n - 4,~~~~n=1,2,3\cdots
\end{equation}
We can also calculate the asymptotic form for the number of distinct sites visited in a level-$n$ walk. We use the well known result that for a generating function of the form
\begin{equation}
\sum_n a_nx^n = \frac{1}{x^{\alpha}}\left[A\left(x\right) + B\left(x\right)\left(1 - x/r\right)^{-\beta}\right]
\label{genseries}
\end{equation}
the asymptotic form for $a_n$ is given by
\begin{equation}
a_n \simeq \frac{B\left(r\right)n^{\beta - 1}}{r^{\alpha}\Gamma\left(\beta\right)r^n}
\end{equation}
where $r$ is the radius of convergence of the series in eq.(\ref{genseries}), and find that for the triangular lattice of three colors, the asymptotic form for the number 
of level-$n$ walks is, from eq.(\ref{c3z3series}) $a_n \simeq \frac{9}{2}n^2$
and for the five colored triangular lattice, this is given from eq.(\ref{c3z5series}) as $a_n \simeq \frac{15}{2}n^2$. 
\begin{figure}[t!]
\begin{minipage}[b]{0.5\linewidth}
\includegraphics[width=2.8in,height=2.3in]{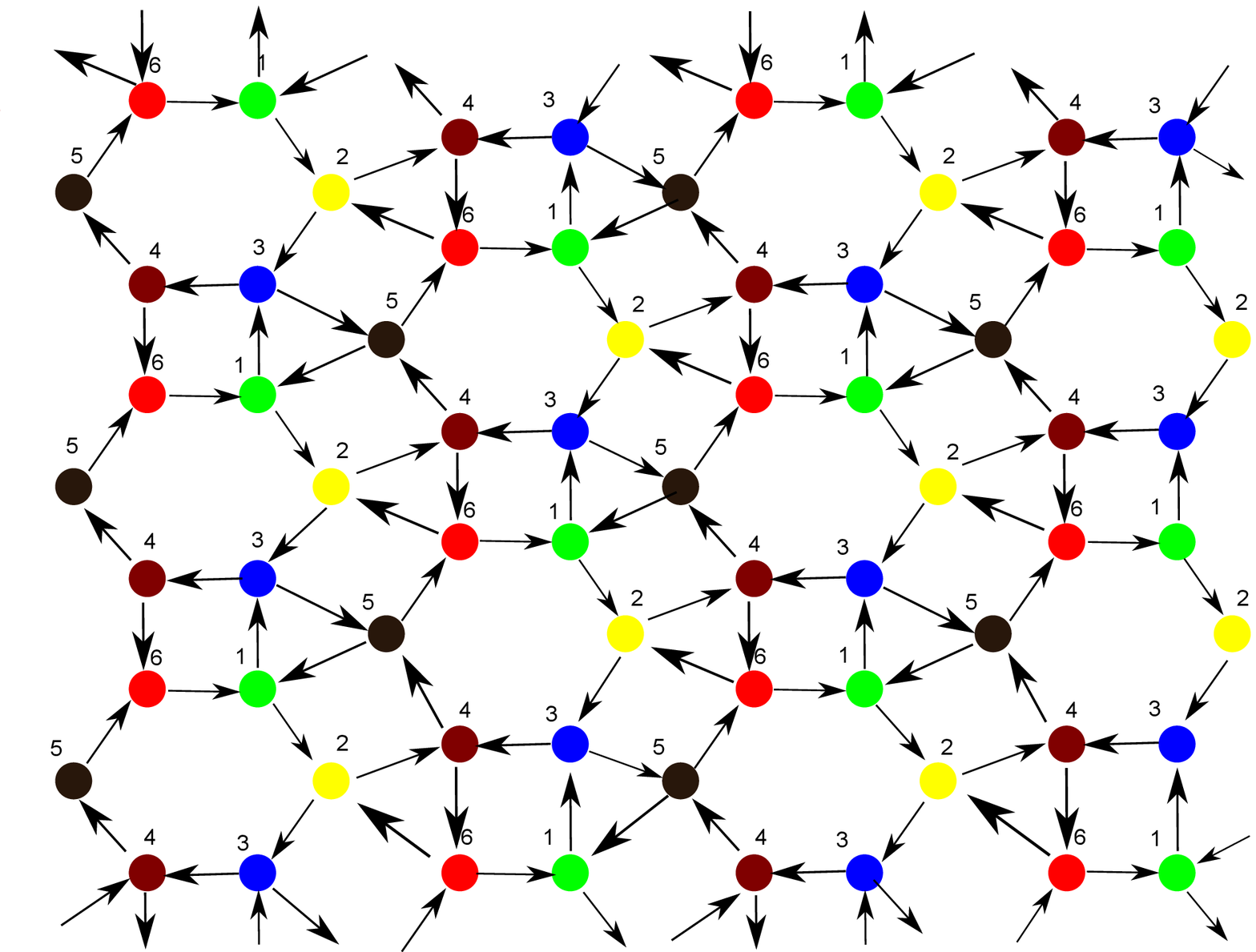}
\caption{An infinite 6-colored directed periodic lattice.}
\label{dP31}
\end{minipage}
\hspace{0.2cm}
\begin{minipage}[b]{0.5\linewidth}
\includegraphics[width=2.8in,height=2.3in]{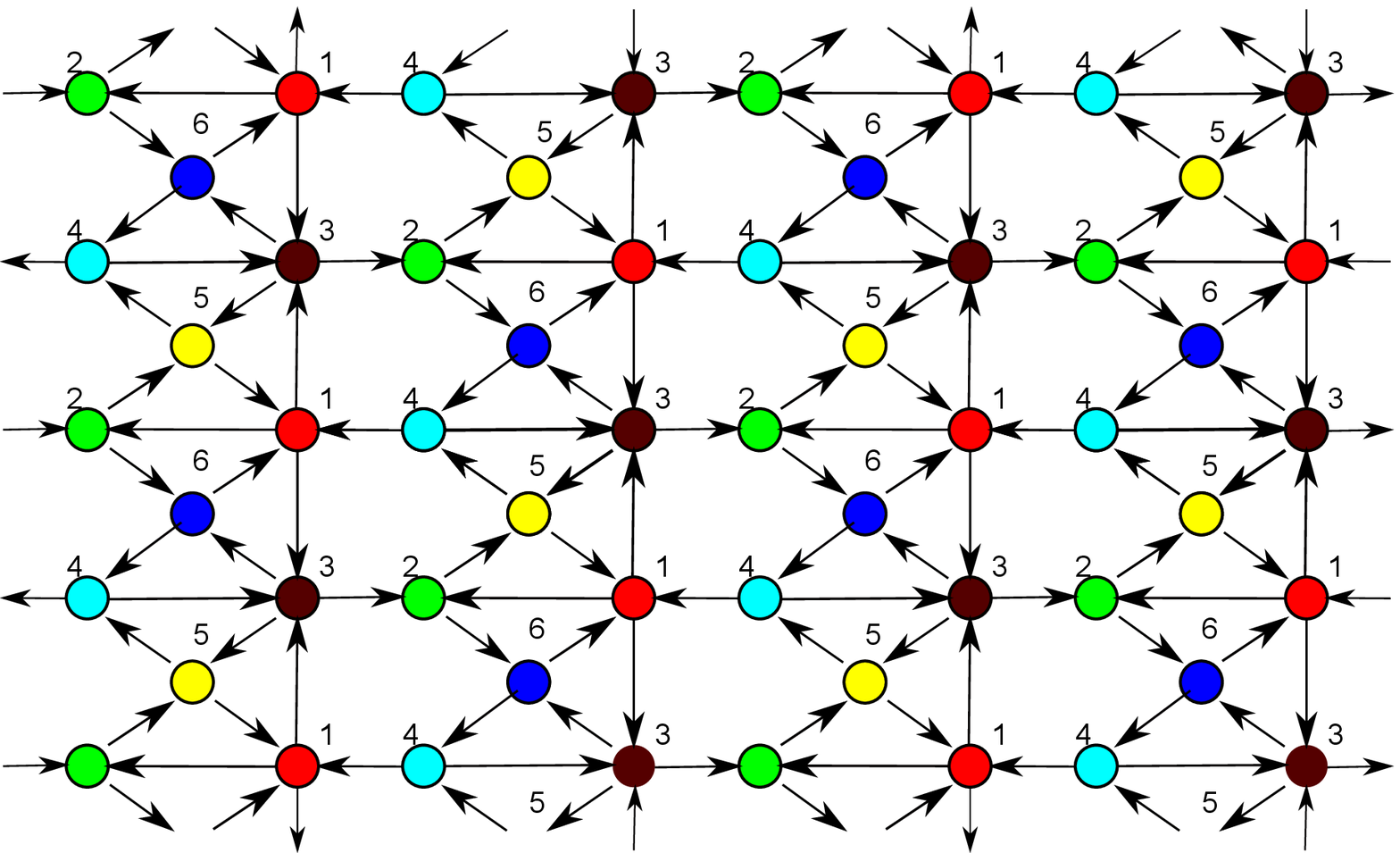}
\caption{Another infinite 6-colored directed periodic lattice.}
\label{dP32}
\end{minipage}
\end{figure}

Although the case of the square and the triangular lattice were simple to deal with, we now show that similar analyses can be carried out for more complicated lattices,
borrowing results that already exist in the string theory literature \cite{hanany}. 
These examples are for D(irichlet)-branes probing (resolutions of) singular spaces that are not orbifolds
(and thus the Molien series is less useful here). 
Consider, for example, the four colored lattice of fig.(\ref{dP1}). The gauge theory here corresponds to D(irichlet)-branes probing 
a resolution of the first del Pezzo surface and the lattice of fig.(\ref{dP1}) is an infinite version of
the quiver diagram for this theory. We simply state the result here for
the number of distinct points visited in a level-$n$ random walk. The generating function for this is given by 
\begin{equation}
{\mathcal F}_4 = \frac{\left(1 + 6x + x^2\right)}{\left(1 - x\right)^3}
\end{equation}
with the asymptotic form of the coefficient being given by $a_n = 4n^2$. The recursion relation for the level-$n$ walk is, in this example,
\begin{equation}
M_n  = M_{n-1} + 8n,~~~~n=1,2,3\cdots
\end{equation}
For the lattices given in figs.(\ref{dP31}) and (\ref{dP32}), the generating function for level-$n$ walks is 
\begin{equation}
{\mathcal F}_5 =  \frac{\left(1 + 4x + x^2\right)}{\left(1 - x\right)^3}
\end{equation}
and the recursion relation is given by
\begin{equation}
M_n = M_{n-1} + 6n,~~~~n=1,2,3\cdots
\end{equation}
Note that for these lattices, walks from one colored site to a site of the same color are possible for all $n$-step walks beginning from $n=3$. The level-$1$ here
consists of all $3$, $4$, $5$ and some $6$ step walks (which lead back to the same color). 
Level $2$ consists of the remaining $6$ step walks, as well as all $7$, $8$ step walks, and some $9$, $10$, 
$11$ and $12$ step walks. Similar considerations holds for the other levels. The lattices for figs.(\ref{dP31}) and (\ref{dP32}) are different, but interestingly, 
they have the same generating function for level-$n$ walks. This is because of the fact that the corresponding operator counting problem in the gauge theories
yield identical results, as these lattices correspond to equivalent gauge theories living on D(irichlet)-branes probing the resolution of the third del Pezzo surface. 

In conclusion, we have shown here that a problem of operator counting in gauge theories inspired from string theory can be recast as a 
random walk problem, involving level-$n$ walks. 
We have provided a few explicit examples here, but emphasize that the procedure holds for an infinite number of examples for which
the operator counting problem in gauge theory has been obtained \cite{hanany}. The asymptotic behaviour of the number of distinct points reached in a level-$n$ walk goes as
$n^2$ for all cases considered here, and this seems to be a generic feature of our analysis. The methods presented here might be useful in trapping problems
with traps placed on sites of particular colors, and the probability of a random walker to be trapped in a level-$n$ walk can be calculated. We leave this for 
a future study.

\begin{center}
\bf{Acknowledgements}
\end{center}
We  sincerely thank V. Subrahmanyam for very useful discussions. SM thanks CSIR India, for financial support through grant no. 09/092(0792)/2011-EMR-I.

 \end{document}